\documentclass{ieeeaccess}
\usepackage{cite}
\usepackage{amsmath}
\usepackage{amssymb}
\usepackage{stfloats}
\usepackage{booktabs}
\usepackage{tabularx}
\usepackage{array}
\newcolumntype{Y}{>{\raggedright\arraybackslash}X}
\usepackage{graphicx}
\usepackage{textcomp}
\usepackage{relsize}
\usepackage{hyperref}
\usepackage[normalem]{ulem}
\usepackage{placeins}
\usepackage{url}

\makeatletter
\def\xfigwd{\textwidth}
\def\appendices{\par%
    \def\theHsection{Appendix.\Alph{section}}%
    \xdef\Hy@chapapp{appendix}%
    \setcounter{section}{-1}%
    \setcounter{subsection}{0}%
    \setcounter{subsubsection}{0}%
    \setcounter{paragraph}{0}%
    \def\thesection{\Alph{section}}%
    \def\thesectiondis{APPENDIX \Alph{section}}%
    \refstepcounter{section}%
    \setcounter{section}{0}%
}
\makeatother
\def\BibTeX{{\rm B\kern-.05em{\sc i\kern-.025em b}\kern-.08em
    T\kern-.1667em\lower.7ex\hbox{E}\kern-.125emX}}
\begin{document}
\vol{XX}
\year{2026}

\title{System-Design Optimization of a Single-Transmitter, Single-Fiber QKD-based ITS Communication Link}
\author{\uppercase{Shivang Srivastava}\authorrefmark{1,2},
\uppercase{Guillaume Ricard}\authorrefmark{1,2},  \uppercase{Yves Jaouën}\authorrefmark{1}, \uppercase{Matteo Schiavon}\authorrefmark{1} and \uppercase{Romain Alléaume}\authorrefmark{1,2}}
\address[1]{LTCI, Télécom Paris, Institut Polytechnique de Paris, 19 Place Marguerite Perey, 91120 Palaiseau, France}
\address[2]{Inria Saclay, 1 Rue Honoré d'Estienne d'Orves, 91120 Palaiseau, France}
\tfootnote{This work was supported by the French National Research Agency (ANR) via the France 2030 program under the PEPR Quantum project QCommTestbed (ANR-22-PETQ-0011).}

\markboth
{Shivang Srivastava \headeretal: System-Design Optimization of a Single-Transmitter, Single-Fiber QKD-based ITS Communication Link}
{Shivang Srivastava \headeretal: System-Design Optimization of a Single-Transmitter, Single-Fiber QKD-based ITS Communication Link}

\corresp{Corresponding author: Shivang Srivastava (email: shivang.srivastava@telecom-paris.fr).}

\begin{abstract}
Deploying QKD-based secure communication at scale requires reducing duplicated optical hardware while preserving the rate of protected user data. Earlier studies established the physical feasibility of carrying CV-QKD and classical signals over shared infrastructure, motivating a quantitative comparison between optical hardware savings and the OTP-encrypted rate. We perform this comparison by defining the secure communication rate (SCR) as the rate jointly supported by secret-key generation and classical transport, and using it to optimize a shared discrete-modulation CV-QKD link. The model combines a binomial 64-QAM security bound with a trusted receiver, protocol traffic, spectral leakage - optical power leaking into the quantum receive band - and Raman scattering. At 25 km, the asymptotic model gives an optimized SCR of 12.079 Mbps, compared with 12.095 Mbps for a separated-link reference, while reducing the transmitter and fiber counts from two to one each in the modeled direction. Increasing channel separation suppresses spectral leakage, but the resulting SCR gain diminishes once receiver bandwidth limits the rate. Under the adopted receiver-noise law, increasing the bandwidth budget from 350 to 750 MHz raises the projected 25-km SCR to 24.535 Mbps despite higher trusted electronic noise. For the studied metropolitan link, sharing saves optical resources with little rate loss, while receiver-bandwidth improvements provide greater benefit than further spectral separation. Together, these results suggest an integration path: share the transmitter and fiber, limit unwanted classical power in the quantum receive band, and increase usable receiver bandwidth where the noise budget permits.
\end{abstract}
\begin{keywords}
Continuous-variable quantum key distribution, information-theoretically secure communication, optical fiber communication, secure communication rate.
\end{keywords}

\titlepgskip=-15pt

\maketitle

\section{Introduction}
\label{sec:introduction}

\PARstart{I}{nformation-theoretically} secure (ITS) communication requires two coupled functions: a mechanism that establishes secret key material with information-theoretic security and an encryption mechanism that uses this material to protect user data. Quantum key distribution (QKD) provides the former through quantum-mechanical principles, while one-time-pad (OTP) encryption provides the latter when the key is uniform, secret, and used only once \cite{BennettBrassard1984,GisinRMP2002}. QKD is therefore one component of a broader secure-communication service. The service must also provide classical transport in both directions for parameter estimation, reconciliation, authentication, and transmission of the OTP-encrypted payload. A positive secret key rate (SKR) becomes operationally useful only when these support tasks and the protected payload can be transported together.

Continuous-variable quantum key distribution (CV-QKD) is attractive for integrated optical secure communication because it uses coherent states, coherent detection, and telecom-compatible components \cite{GrosshansGrangier2002,GrosshansNature2003,DiamantiLeverrier2015}. The first mature family of CV-QKD protocols was based on Gaussian-modulated coherent states and reverse reconciliation, which enabled operation through lossy channels and established the standard transmittance and excess-noise parameter-estimation viewpoint \cite{GrosshansGrangier2002,GrosshansNature2003,Grosshans2005PRL}. The associated security theory has been developed through covariance-matrix methods, Holevo-information bounds, composable finite-size proofs, and careful treatment of calibration and trusted receiver noise \cite{DiamantiLeverrier2015,Leverrier2015Composable,Jain2022Composable}. Experimentally, Gaussian-modulated CV-QKD has demonstrated high-rate and field-compatible operation using coherent receivers and modern reconciliation schemes \cite{Jouguet2014,CVQKD_25km_Weerasinghe2023}.

Discrete-modulation CV-QKD (DM-CV-QKD) has become increasingly relevant for system design because its constellations are close to classical digital modulation formats and can be implemented with standard digital-to-analog conversion and optical in-phase/quadrature (IQ) modulation hardware \cite{LeverrierGrangier2009,Ghorai2019PRX}. Security analyses for discrete modulation have progressed from asymptotic approaches to finite-size and coherent-attack frameworks \cite{Ghorai2019PRX,Kanitschar2023PRXQuantum,Baeuml2024Quantum,PascualGarcia2025}. In parallel, high-rate DM-CV-QKD experiments and recent composable finite-size demonstrations indicate that discrete modulation is becoming a practical candidate for deployment-oriented CV-QKD \cite{PanHighRate2022,Hajomer2025Composable}.

QKD-secured communication has already been demonstrated at the service level in several forms. Mink et al.\ showed real-time video protected by OTP keys generated by a high-speed QKD test bed, while the Tokyo QKD Network demonstrated 128~kbit/s secure television (TV) conferencing using OTP keys over a trusted-relay QKD network \cite{Mink2006,Sasaki2011}. Jouguet et al.\ noted in their SEQURE field test that sub-Mbit/s QKD secret-key rates favor frequent Advanced Encryption Standard (AES) key renewal for practical Gbit/s-class applications \cite{jouguet2012field}. Choi et al.\ showed QKD-assisted key refresh for high-rate AES-encrypted data traffic over installed fiber, and the Defense Advanced Research Projects Agency (DARPA) Quantum Network demonstrated QKD-supported network-security services \cite{choi2014field,Elliott2003}. Together, these demonstrations connect QKD key generation to an encrypted communication service. In the OTP setting considered here, the corresponding service metric is the payload rate jointly supported by key generation and classical transport.

We treat the classical and DM-CV-QKD layers as a coupled system running over shared hardware. The classical layer makes the generated keys operationally useful by carrying parameter-estimation disclosure, reconciliation, authentication, and the OTP-encrypted payload. At the same time, imperfect isolation, spectral overlap, and receiver nonidealities allow the classical signal to impair the quantum channel. The QKD protocol consumes classical capacity, while the classical channel contributes noise to the quantum channel. The useful encrypted payload rate is consequently set by the joint feasibility of secret-key generation and classical transport \cite{Almeida2024IET}.

A related body of work has studied the co-propagation of QKD and classical optical channels, primarily in wavelength-division-multiplexed systems where a weak quantum signal travels alongside strong classical traffic \cite{Kumar2015,Hajomer2025Coexist}. More integrated approaches have demonstrated simultaneous classical communication and CV-QKD using shared transmitter hardware and coherent infrastructure \cite{Qi2016,Shao2025}. These results establish the physical feasibility of sharing optical resources and motivate a system-level study of the encrypted payload that such an architecture can support.

The modeled architecture uses one transmitter chain and one standard single-mode fiber for the classical and DM-CV-QKD signals in the Alice-to-Bob direction. A two-way service can use a second fiber with a corresponding shared transmitter at Bob for Bob-to-Alice traffic. Reverse reconciliation for each quantum transmission then travels over the fiber in the opposite direction. Polarization and frequency multiplexing provide channel separation within each direction, while the present platform retains dedicated receiver and digital signal-processing (DSP) chains. The reference architecture builds on the shared quantum/classical transmission studies in \cite{RicardThesis2025, AymericThesis2022}, with selected characterization data taken from \cite{RicardThesis2025}. The numerical model combines the binomial 64-QAM covariance bound with a trusted receiver, physical leakage and Raman terms, the classical-communication overhead of QKD, and an outer SCR optimization for the Alice-to-Bob direction.

To express this coupling at the service level, we define the secure communication rate (SCR) as the OTP-encrypted payload rate supported simultaneously by secret-key generation and the available classical transport. With $R_c^{\mathrm{net}}$ the net forward classical throughput and $R_{\mathrm{PE}}$ the parameter-estimation disclosure rate, a feasible configuration satisfies
\begin{equation}
R_c^{\mathrm{net}}-R_{\mathrm{PE}} \geq \mathrm{SCR}.
\label{eq:classical-scr-constraint}
\end{equation}

Reverse-reconciliation traffic travels from Bob to Alice on the opposite fiber, whose net capacity is provisioned separately in the one-direction model described in Section~\ref{sec:architecture}. This definition captures three forms of coupling. The QKD protocol consumes classical capacity through parameter-estimation and reconciliation traffic; the classical signal contributes leakage and Raman noise to the quantum channel; and both layers share finite spectral, receiver-bandwidth, and launch-power resources.

The contributions of this study are threefold. First, we formulate SCR as the OTP-encrypted payload jointly supported by key generation and classical transport. Second, we place binomial 64-QAM security, the trusted receiver, sharing-induced noise, support traffic, and hardware accounting within a common physical model. Third, we jointly optimize the quantum and classical symbol rates, center-frequency separation, modulation variance, classical launch power, and fiber length, and examine receiver bandwidth together with its electronic-noise contribution. The resulting sweeps identify how spectral separation, receiver design, and distance-dependent Raman scattering determine the sharing penalty. Before developing this model, Section~\ref{sec:related} positions the proposed architecture within earlier QKD-secured communication systems and establishes the resource comparison used throughout the paper.

\section{Architecture-Level Positioning and Resource Comparison}
\label{sec:related}

Table~\ref{tab:qkd_encrypted_service_comparison} provides an architecture-level view of the sharing benefit. It reconstructs the transmitter, fiber, and receiver/DSP resources of representative QKD-secured communication systems and places their reported service rates alongside the corresponding equal-weight hardware index.

\begin{table*}[!t]
\centering
\caption{Partial hardware resource comparison of QKD-secured communication systems. \(N_{\mathrm{Tx}}\), \(N_{\mathrm{fib}}\), and \(N_{\mathrm{Rx/DSP}}\) denote reconstructed transmitter-chain, fiber, and receiver/DSP resources. The equal-weight hardware index is \(C_{\mathrm{hw}}^{(1)}=N_{\mathrm{Tx}}+N_{\mathrm{fib}}+N_{\mathrm{Rx/DSP}}\). The reported rate is the protected OTP payload for OTP systems and the QKD key-generation rate available for AES key refresh or standalone QKD operation; the corresponding protocols and security assumptions differ. The count for this work describes one modeled transmission direction; the opposite fiber and its shared transmitter are discussed separately in Section~\ref{sec:architecture}. The entry for this work is the simulated platform-rate SCR evaluated at \(L=25~\mathrm{km}\), \(f_c=4~\mathrm{GBd}\), \(f_q=125~\mathrm{MBd}\), and \(\Delta f=3~\mathrm{GHz}\), with \(P_c\) set to the minimum value satisfying the classical signal-to-noise-ratio
(SNR) constraint and \(V_A\) optimized. Its absolute-power scale follows the assumed $P_{c,0}=1~\mathrm{mW}$ reference described in Section~\ref{subsec:reference_configuration}. The optimized \(269.23~\mathrm{MBd}\) design projection is reported separately in Section~\ref{sec:codesign_study}.}
\label{tab:qkd_encrypted_service_comparison}
\begingroup
\scriptsize
\setlength{\tabcolsep}{4.1pt}
\renewcommand{\arraystretch}{1.16}
\begin{tabular}{>{\raggedright\arraybackslash}p{2.25cm}
                >{\raggedright\arraybackslash}p{2.05cm}
                c c c c r r}
\toprule
\textbf{Work} &
\textbf{Security Service} &
\(\boldsymbol{N_{\mathrm{Tx}}}\) &
\(\boldsymbol{N_{\mathrm{fib}}}\) &
\(\boldsymbol{N_{\mathrm{Rx/DSP}}}\) &
\(\boldsymbol{C_{\mathrm{hw}}^{(1)}}\) & 
\textbf{Reported Rate} & \textbf{Hardware-Normalized Rate}
 \\
& & & & & & \textbf{(kbit/s)} & \(\boldsymbol{\mathrm{Rate}/C_{\mathrm{hw}}^{(1)}}\) \\
\midrule

Mink \emph{et al.}~\cite{Mink2006}
& OTP
& 2 & 2 & 2 & 6
& \(\sim\)1{,}000 & \(\sim\)170 \\

Sasaki \emph{et al.}~\cite{Sasaki2011}
& OTP
& 3 & 4 & 3 & 10
& 128 & 13 \\

Jouguet \emph{et al.}~\cite{jouguet2012field}
& AES
& 2 & 2 & 2 & 6
& \(\sim\)0.6 & \(\sim\)0.1 \\

Eraerds \emph{et al.}~\cite{eraerds2010singlefibre}
& AES
& 2 & 1 & 2 & 5
& 0.511 & 0.10 \\

Choi \emph{et al.}~\cite{choi2014field}
& AES
& 2 & 1 & 2 & 5
& 160 & 32 \\

Dynes \emph{et al.}~\cite{dynes2016ultrahigh}
& AES
& 2 & 2 & 2 & 6
& 1{,}900 & \(\sim\)320 \\

Jain \emph{et al.}~\cite{jain2024future}
& AES
& 2 & 1 & 2 & 5
& 434.8 & 87 \\

Toshiba MU~\cite{toshibaMU}
& QKD
& 2 & 1 & 2 & 5
& 300 & 60 \\

\midrule

\textbf{This work}\newline(simulated, asymptotic)
& \textbf{OTP}
& \textbf{1} & \textbf{1} & \textbf{2} & \textbf{4}
& \textbf{5{,}523} & \(\boldsymbol{\sim}\)\textbf{1{,}381} \\

\bottomrule
\end{tabular}
\endgroup
\end{table*}

Under this bookkeeping convention, the simulated platform-rate configuration illustrates the reduction in transmitter and fiber resources for one modeled direction. A symmetric two-way deployment doubles the hardware counts. The comparison is an architecture-level indicator: the protocol, security treatment, and service definition associated with each rate remain those of the cited system. Section~\ref{sec:codesign_study} reports the jointly optimized shared design separately.

This comparison motivates the integration path in Fig.~\ref{fig:resource_sharing_vision}, which distinguishes the transmitter and fiber resources shared by the proposed architecture from the receiver/DSP resources that remain separate in the current platform.

\section{Shared-Hardware Integrated ITS Architecture}
\label{sec:architecture}

\begin{figure*}[!b]
\centering
\includegraphics[width=\textwidth]{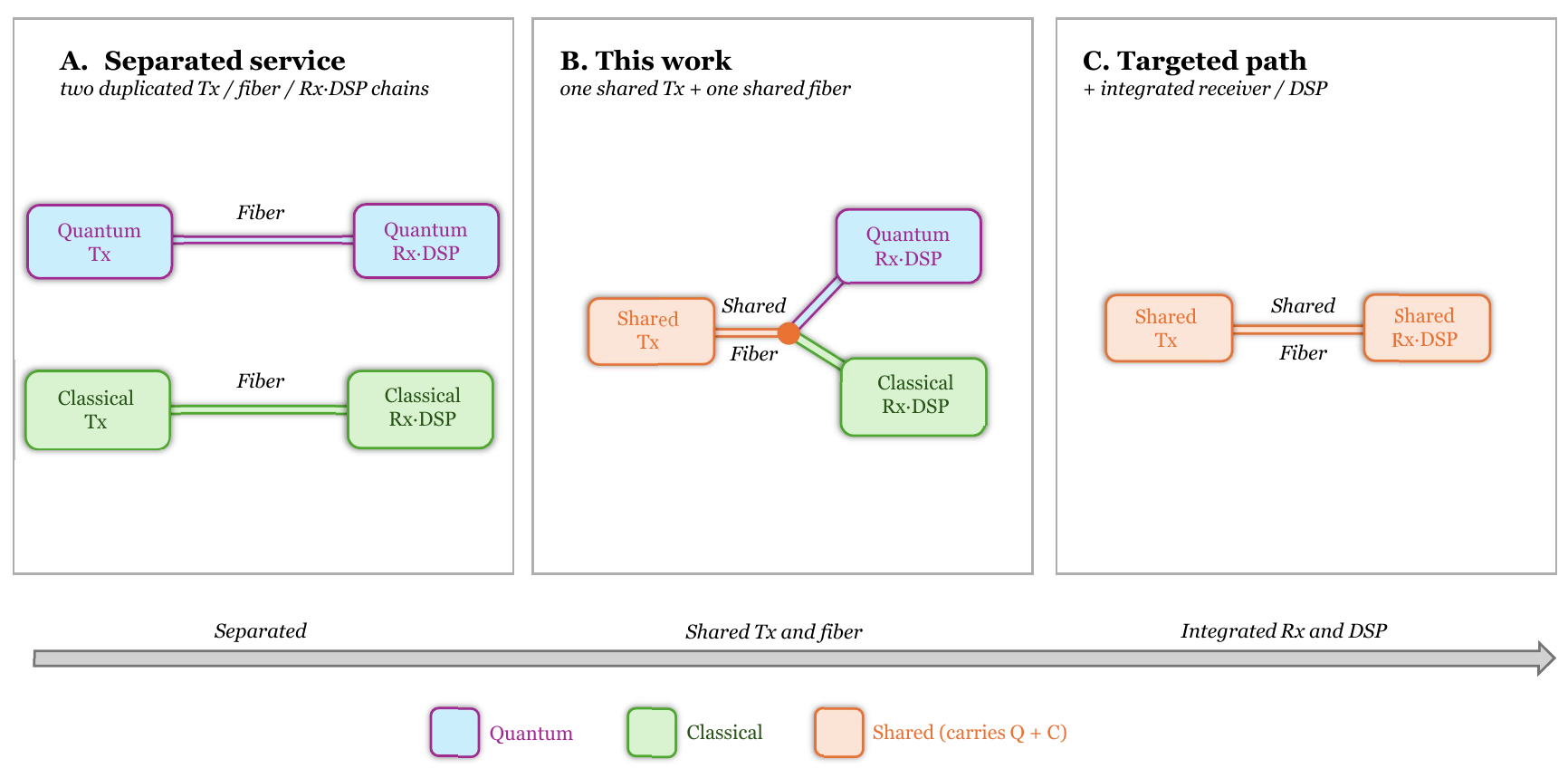}
\caption{\parbox{0.90\textwidth}{\raggedright Hardware resource motivation per transmission direction for the proposed shared classical/quantum architecture. A separated implementation duplicates transmitter (Tx), fiber, receiver (Rx), and DSP resources. The proposed implementation shares the Tx and fiber within each direction while retaining separate Rx/DSP chains in the current platform. A two-way service places one shared Tx at each end and uses two fibers, one per transmission direction. The integration path extends to shared Rx/DSP resources when the associated SCR penalty remains small.}}
\label{fig:resource_sharing_vision}
\end{figure*}

\begin{figure*}[!t]
\centering
\includegraphics[width=0.95\textwidth]{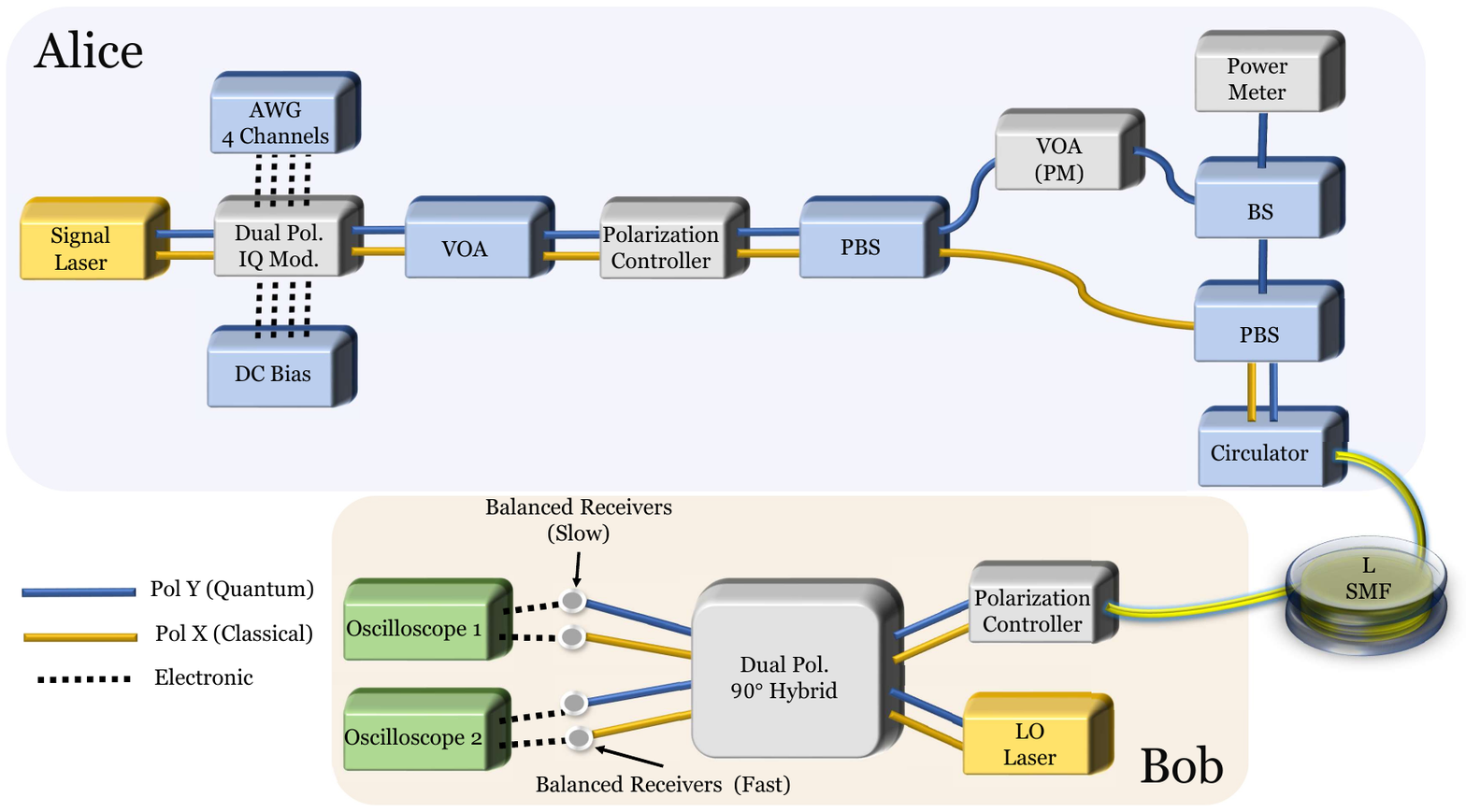}

\smallskip
{\small Two-way deployment: Bob's shared Tx
$\longrightarrow$ opposite fiber $\longrightarrow$ Alice's receivers.}
\caption{\parbox{0.90\textwidth}{\raggedright Reference architecture of the characterized single-transmitter, single-fiber Alice-to-Bob path used in the simulations. Alice generates the quantum and classical signals with a shared transmitter chain and launches them into the same fiber of length $L$. This modeled direction is counted as $N_{\mathrm{Tx}}=N_{\mathrm{fib}}=1$ in Table~\ref{tab:qkd_encrypted_service_comparison}; Bob's quantum and classical receiver/DSP chains remain separate and are counted as $N_{\mathrm{Rx/DSP}}=2$. In a two-way deployment, Bob's shared transmitter would use the opposite fiber for Bob-to-Alice quantum/classical transmission, including the reconciliation data for Alice's quantum signal. The model assumes an independently provisioned net capacity $R_{\rm back}=10~\mathrm{Gbit/s}$ for that return direction.}}
\label{fig:setup}
\end{figure*}

Here, ``reference platform'' denotes the shared classical/CV-QKD
architecture described in \cite{RicardThesis2025}, which provides the
architectural basis and selected parameters for the numerical model.
We specify which hardware components are shared and which remain
dedicated. The middle configuration in Fig.~\ref{fig:resource_sharing_vision} represents one transmission direction: the transmitter and fiber are shared, whereas the receiver and DSP chains remain separate. In a two-way service, Alice and Bob each have a shared transmitter and use one fiber to transmit toward the other end. In the Alice-to-Bob platform shown in Fig.~\ref{fig:setup}, one optical source, one dual-polarization IQ modulator, and one arbitrary-waveform-generator (AWG) chain generate the classical and quantum signals, which share a launch fiber and are separated through polarization and frequency multiplexing. The classical launch power $P_c$, center-frequency separation $\Delta f$, quantum modulation variance $V_A$, and channel length $L$ then connect the physical architecture to the excess-noise budget, QKD classical-overhead rate, and key-generation feasibility.

Fig.~\ref{fig:setup} shows the shared Alice-to-Bob optical path. Parameter-estimation disclosure and OTP ciphertext travel from Alice to Bob on its classical channel. Bob's reverse-reconciliation data travel back to Alice on the opposite fiber. In a two-way deployment, Bob's shared transmitter can also send a quantum signal, its own parameter-estimation data, and protected payload on that fiber; Alice's reconciliation data would then travel on the Alice-to-Bob fiber. Both classical directions must be authenticated, while reconciliation data can be public \cite{GisinRMP2002}. The two fibers isolate co-propagating Raman contributions between directions. Within Bob's fiber, however, the classical power needed for return traffic would affect Bob's quantum signal and must be included in a two-way optimization.

The numerical study optimizes the Alice-to-Bob shared link with an independently provisioned net return capacity $R_{\rm back}=10~\mathrm{Gbit/s}$ for reconciliation. The single-transmitter, single-fiber designation and the index in Table~\ref{tab:qkd_encrypted_service_comparison} apply to the modeled direction.

The resource motivation can be expressed through the hardware count:
\begin{equation}
C_{\mathrm{hw}}
=
\omega_{\mathrm{tx}}N_{\mathrm{Tx}}
+
\omega_{\mathrm{fib}}N_{\mathrm{fib}}
+
\omega_{\mathrm{rx}}N_{\mathrm{Rx/DSP}}
+
\omega_{\mathrm{cal}}N_{\mathrm{cal}},
\label{eq:hardware_resource_count}
\end{equation}
where $N_{\mathrm{Tx}}$, $N_{\mathrm{fib}}$, $N_{\mathrm{Rx/DSP}}$, and
$N_{\mathrm{cal}}$ denote transmitter chains, fiber resources, receiver/DSP chains, and calibration/control loops, respectively. The weights $\omega_i$ represent the deployment-dependent importance assigned to each resource. Because these weights and the calibration/control-loop counts vary among demonstrations, the comparison also uses the equal-weight partial index
\begin{equation}
C_{\mathrm{hw}}^{(1)}
=
N_{\mathrm{Tx}}
+
N_{\mathrm{fib}}
+
N_{\mathrm{Rx/DSP}},
\label{eq:hardware_resource_index}
\end{equation}
as a transparent equal-weight resource index.

In each transmission direction, the reduction lies in $N_{\mathrm{Tx}}$ and $N_{\mathrm{fib}}$, because one transmitter chain generates the classical and DM-CV-QKD signals and one fiber carries them. The receiver/DSP chains remain separate; their future integration would further reduce $N_{\mathrm{Rx/DSP}}$. Two such directions would use two shared transmitters and two fibers in total, averaging one of each per direction.

Sharing also couples the physical constraints of the two channels. Increasing $P_c$ improves the classical margin but can raise leakage, Raman noise, relative-intensity-noise transfer, or receiver saturation. Increasing $f_q$ raises the available secret-key rate, while requiring more receiver and analog-to-digital-converter (ADC) bandwidth and tighter phase recovery. These opposing trends make SCR, rather than hardware count alone, the appropriate system objective. Section~\ref{sec:optimizer_model} formalizes this coupling.

\section{SCR Optimization Model}
\label{sec:optimizer_model}

We determine the largest OTP-encrypted payload supported by a chosen hardware configuration and fiber length by combining the service constraint, spectral and noise models, trusted-receiver key rate, and nested optimization. The model uses binomial 64-QAM discrete modulation, heterodyne detection, reverse reconciliation, and a trusted receiver with the parameters specified in Table~\ref{tab:platform_parameters}. The secret fraction is an asymptotic lower bound against collective attacks, while the parameter-estimation fraction enters both the secret-key rate and the classical-overhead allocation. All noise quantities are normalized in shot-noise units (SNU), with their reference planes stated explicitly. Table~\ref{tab:nomenclature} collects
the principal symbols used in Sections~IV and V.

\begin{table*}[!t]
\caption{Nomenclature for the physical and optimization model.}
\label{tab:nomenclature}
\centering
\small
\begin{tabularx}{\textwidth}{@{}lYl@{}}
\toprule
Symbol & Definition & Unit \\
\midrule
$L$, $P_c$ & Fiber length and classical launch power & km, mW \\
$f_c$, $f_q$, $\Delta f$ & Classical and quantum symbol rates and center-frequency separation & GBd, GBd, GHz \\
$B_c$, $B_q$ & Roll-off-inclusive classical and quantum occupied bandwidths & GHz \\
$\rho_c$, $\rho_q$ & Classical and quantum RRC roll-off factors & 1 \\
$B_{\rm tot}$, $B_{\rm guard}$ & Available equivalent-baseband bandwidth and minimum guard band & GHz \\
$V_A$, $T_{\rm ch}$ & Alice modulation variance and optical-channel power transmittance & SNU, 1 \\
$B_{\rm RX}$, $\eta_B$ & Usable two-sided complex-baseband receiver bandwidth and trusted efficiency & GHz, 1 \\
$S_{\rm el}$, $S_{\rm SN}$; $V_{\rm el}$, $V_{\rm SN}$ & Electronic- and shot-noise PSDs; corresponding band-integrated variances & variance/GHz; variance \\
$N_0$, $\xi_{\rm rec}(B_{\rm RX})$ & Shot-noise variance and bandwidth-dependent trusted electronic noise & SNU \\
$\xi_0$, $\kappa_{\rm RX}$ & Receiver-noise intercept and quadratic coefficient & SNU, SNU/GHz$^2$ \\
$\xi_A$, $\xi_{\rm leak}$, $\xi_{\rm Raman}$ & Baseline, leakage-induced, and Raman-scattering input-referred excess noise & SNU \\
$\xi_{\rm link}$ & Total input-referred link excess noise & SNU \\
$\mathbf X_B$, $\lambda_{\rm EC}$ & Joint quantized heterodyne outcome and reconciliation transport allowance & 1, bit/symbol \\
$I(X;Y)$, $I(X;\mathbf X_B)$, $\chi(E;Y)$ & Continuous-output, quantized-output, and Eve--Bob information quantities & bit/symbol \\
$K_q$, $R_c^{\rm net}$, $R_{\rm sup}$ & Secret-key rate, net classical data rate, and QKD classical-overhead rate & Mbit/s \\
$\beta$, $\nu_{\rm PE}$, $b_q$ & Reconciliation efficiency, parameter-estimation fraction, and bits per disclosed quantum symbol & 1, 1, bit/symbol \\
$R_{\rm PE}$, $R_{\rm EC}$, $R_{\rm back}$ & Disclosure rate, reverse reconciliation rate, and assumed net opposite-direction capacity & Mbit/s \\
$\eta_c$, $f_{\rm EC}$ & Net forward transport efficiency and reconciliation provisioning factor & 1 \\
$\epsilon_{\rm OOB}$ & Additional filter-admitted classical power fraction before polarization suppression & 1 \\
$s$, $\mathrm{SCR}$ & Feasible OTP payload rate and its optimized value & Mbit/s \\
$\mathcal{P}_{\rm sh}$ & Shared-link rate penalty & dB \\
\bottomrule
\end{tabularx}
\end{table*}

\subsection{Service rate and support traffic}
\label{subsec:classical_support_models}

For the modeled Alice-to-Bob direction, the encrypted payload is limited by the secret-key rate and the classical throughput left after parameter estimation. Bob-to-Alice reconciliation must fit the separately provisioned opposite-direction capacity. With the overhead rates defined below, the service constraints for this one-direction calculation are
\begin{equation}
 \begin{aligned}
 s&\leq K_q,\qquad s+R_{\rm PE}\leq R_c^{\rm net},\\
 R_{\rm EC}&\leq R_{\rm back}.
 \end{aligned}
 \label{eq:scr_service_constraints}
\end{equation}
At a physical operating point with $R_{\rm EC}\leq R_{\rm back}$ and $R_{\rm PE}\leq R_c^{\rm net}$,
\begin{equation}
 \mathrm{SCR}
 =\left[\min\!\left(K_q,\,R_c^{\rm net}-R_{\rm PE}\right)\right]_+.
 \label{eq:scr_definition}
\end{equation}
Here $[u]_+=\max(u,0)$; operating points that violate the assumed opposite-direction capacity are infeasible. In a simultaneous two-way service, each shared classical channel must also carry reconciliation for the quantum signal transmitted in the other direction. With subscripts $AB$ and $BA$ denoting Alice-to-Bob and Bob-to-Alice quantum transmissions, respectively, the coupled transport conditions are
\begin{align}
 s_{AB}+R_{{\rm PE},AB}+R_{{\rm EC},BA}&\leq R_{c,AB}^{\rm net},\notag\\
 s_{BA}+R_{{\rm PE},BA}+R_{{\rm EC},AB}&\leq R_{c,BA}^{\rm net}.
 \label{eq:two_way_transport_constraints}
\end{align}
Here $R_{{\rm EC},AB}$ is Bob's reconciliation data for Alice's quantum transmission, carried on the Bob-to-Alice fiber; $R_{{\rm EC},BA}$ is Alice's reconciliation data for Bob's quantum transmission, carried on the Alice-to-Bob fiber. Equation~\eqref{eq:two_way_transport_constraints} describes the two-way deployment; the reported one-direction optimizations use Eq.~\eqref{eq:scr_service_constraints} with independent $R_{\rm back}$. For quadrature phase-shift keying (QPSK), the net throughput of the modeled Alice-to-Bob classical channel is
\begin{equation}
 R_c^{\rm net}=2\times10^3\eta_c f_c\quad\mathrm{Mbit/s},
 \label{eq:classical_net_rate}
\end{equation}
where $10^3$ converts GBd to Mbaud and $\eta_c$ accounts for coding, framing, and transport overhead. We use $\eta_c=0.95$ as a transport-efficiency assumption and examine values between $0.90$ and $1.00$. The return capacity $R_{\rm back}$ is specified independently as a net transport rate after its own overhead.

Let $X$ denote Alice's joint 64-QAM symbol, $Y_I$ and $Y_Q$ Bob's two
pre-quantization heterodyne quadratures, and $Q(\cdot)$ the same six-bit
uniform scalar quantizer
on each quadrature. Bob's joint discrete outcome is
\begin{equation}
 \mathbf X_B=\bigl(Q(Y_I),Q(Y_Q)\bigr).
 \label{eq:joint_quantized_heterodyne_outcome}
\end{equation}
For each quadrature, the quantizer covers the symmetric interval
$[-4\sigma_B,4\sigma_B]$, where $\sigma_B$ is the corresponding
pre-quantization standard deviation. Its cell width is
$\Delta_q=8\sigma_B/64$; the $63$ finite internal boundaries are
$-4\sigma_B+m\Delta_q$, $m=1,\ldots,63$, and the two outer cells absorb
the tails.
We use this six-bit quantizer to dimension the reconciliation traffic in the numerical model.
The parameter-estimation disclosure uses
$b_q=\log_2|\mathcal{X}_q|=\log_2 64=6$ bits per revealed quantum symbol,
where $\mathcal{X}_q$ is the quantum constellation. Because square 64-QAM
and the equivalent discrete channel factorize into independent, identically distributed
in-phase and quadrature components,
\begin{equation}
 \begin{aligned}
 H(\mathbf X_B)&=2H\bigl(Q(Y_I)\bigr),\\
 I(X;\mathbf X_B)&=2I\bigl(X_I;Q(Y_I)\bigr).
 \end{aligned}
 \label{eq:quantized_information_factorization}
\end{equation}
Both quantities are evaluated from the exact quantizer-cell probabilities defined in Eq.~\eqref{eq:quantized_cell_probabilities}. The Slepian--Wolf limit on the reconciliation disclosure is
$H(\mathbf X_B|X)=H(\mathbf X_B)-I(X;\mathbf X_B)$
\cite{CoverThomas2006}. For classical-link provisioning we use
\begin{equation}
 \lambda_{\rm EC}
 =f_{\rm EC}\left[H(\mathbf X_B)-I(X;\mathbf X_B)\right]_+,
 \qquad f_{\rm EC}=1/0.95.
 \label{eq:reconciliation_disclosure}
\end{equation}
The factor $f_{\rm EC}$ sets a conservative transport allowance following the support-traffic model in \cite{RicardThesis2025}, independently of the reconciliation efficiency $\beta=0.95$ in Eq.~\eqref{eq:secret_fraction}. The six-bit quantizer determines the return-traffic budget $\lambda_{\rm EC}$, while the $\beta I(X;Y)$ term in the key-rate expression assumes reconciliation that exploits Bob's soft information; an implementation must achieve the assumed efficiency within this budget and account for its actual information disclosure in the security analysis.
The total QKD classical-overhead rate is therefore
\begin{equation}
 R_{\rm sup}
 =10^3 f_q\!\left[\nu_{\rm PE}b_q+(1-\nu_{\rm PE})\lambda_{\rm EC}\right]
 +R_{\rm auth}.
 \label{eq:support_rate}
\end{equation}
The directional rates are
\begin{equation}
 \begin{aligned}
 R_{\rm PE}&=10^3f_q\nu_{\rm PE}b_q,\\
 R_{\rm EC}&=10^3f_q(1-\nu_{\rm PE})\lambda_{\rm EC}.
 \end{aligned}
 \label{eq:directional_support_rates}
\end{equation}
Here $\nu_{\rm PE}$ is the fraction of quantum symbols disclosed for parameter estimation. The reported calculations set the additional fixed allowance $R_{\rm auth}=0$, with transport overhead represented by $\eta_c$ and the net return capacity. Any separately specified authentication/control load must be assigned to its direction in Eq.~\eqref{eq:scr_service_constraints}.

\subsection{Spectral occupancy and shared-link noise}
\label{subsec:shared_noise_model}

\begin{figure*}[!t]
\centering
\includegraphics[width=0.94\textwidth]{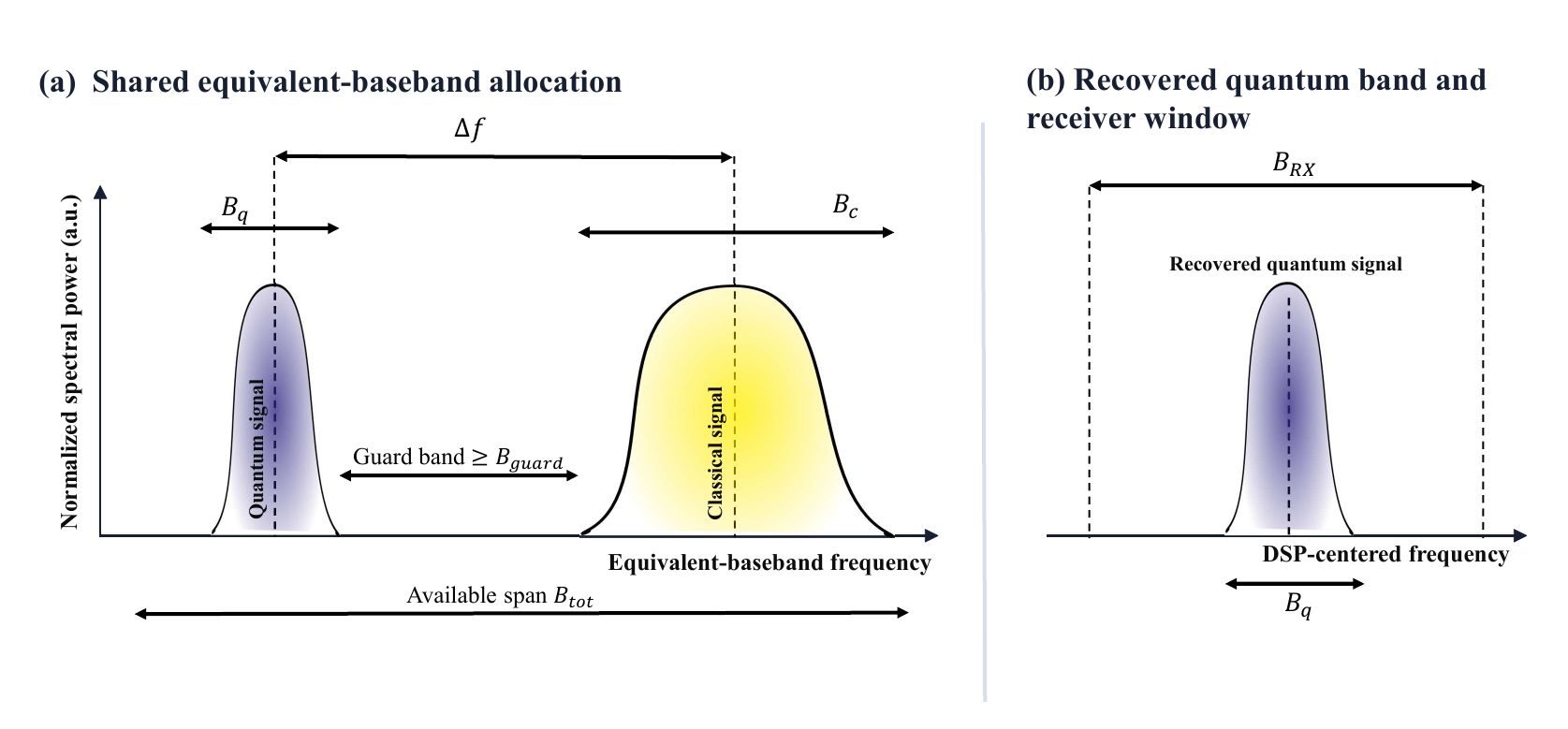}
\caption{\parbox{0.92\textwidth}{\raggedright
Schematic frequency-domain definitions used by the model. (a) Normalized
spectral-power envelopes of the quantum and classical signals, with
roll-off-inclusive occupied bandwidths $B_q$ and $B_c$, center-frequency
separation $\Delta f$, minimum edge-to-edge guard band $B_{\rm guard}$,
and available equivalent-baseband bandwidth $B_{\rm tot}$. (b) After
heterodyne downconversion and DSP centering, the recovered quantum band of
width $B_q$ lies within the modeled usable receiver bandwidth $B_{\rm RX}$.
The envelopes and receiver window are schematic representations of spectral placement, occupied bandwidth, and the adopted receiver budget; amplitudes are in arbitrary units and widths are not drawn to scale.}}
\label{fig:frequency_layout}
\end{figure*}

The shared transmitter maps the classical and quantum waveforms onto frequency-separated bands around a common optical carrier. In the equivalent-baseband representation used for the spectral constraints, their layout determines both spectral feasibility and the classical leakage admitted by the quantum-channel receive filter. For root-raised-cosine (RRC) pulse shaping, the roll-off-inclusive occupied bandwidths are
\begin{equation}
 B_c=(1+\rho_c)f_c,
 \qquad
 B_q=(1+\rho_q)f_q.
 \label{eq:occupied_bandwidths}
\end{equation}
Here $\rho_c$ and $\rho_q$ are the classical and quantum RRC roll-off
factors, respectively.
The center-frequency separation must provide the sum of the two half-bandwidths and the minimum edge-to-edge guard band,
\begin{equation}
 \Delta f\geq \frac{B_c+B_q}{2}+B_{\rm guard}.
 \label{eq:inner_spectral_constraint}
\end{equation}
The available equivalent-baseband and quantum-receiver bandwidth constraints are
\begin{equation}
 \Delta f+\frac{B_c+B_q}{2}\leq B_{\rm tot},
 \qquad B_q\leq B_{\rm RX}.
 \label{eq:outer_receiver_constraints}
\end{equation}
Here $B_{\rm tot}$ is the equivalent-baseband bandwidth available to the shared
transmitter. The occupied span already includes the interchannel separation; $B_{\rm guard}$ is applied only between the two bands in Eq.~\eqref{eq:inner_spectral_constraint}. In the optimization, $B_{\rm RX}$ denotes the combined detector/ADC/DSP two-sided complex-baseband occupied-bandwidth budget after heterodyne downconversion and DSP centering, represented by the interval $[-B_{\rm RX}/2,B_{\rm RX}/2]$, with feasibility requiring $B_q\leq B_{\rm RX}$. We adopt $B_{\rm RX}=0.35~\mathrm{GHz}$ as a modeling budget informed by the receiver chain.
Figure~\ref{fig:frequency_layout} summarizes the
frequency-domain convention and the associated notation.

Let $S_c(f)$ be the normalized classical-signal power spectral density (PSD),
with $\int_{-\infty}^{\infty}S_c(f)\,df=1$, and let $H_q(f)$ be the frequency
response of the quantum-channel receive filter. The fraction of classical
power passed by this filter is
\begin{equation}
 \Omega(\Delta f)
 =\int_{-\infty}^{\infty}S_c(f-\Delta f)|H_q(f)|^2\,df.
 \label{eq:spectral_overlap}
\end{equation}
Equation~\eqref{eq:spectral_overlap} is the standard output-power relation for a linear time-invariant (LTI) filter applied to a stationary input spectrum \cite{Oppenheim1997}. With polarization extinction ratio (PER) and $\eta_{\rm pol}=10^{-\mathrm{PER}/10}$, the input-referred leakage excess noise is
\begin{equation}
 \xi_{\rm leak}
 =\frac{2\eta_{\rm pol}P_c\Omega(\Delta f)}{h\nu_q f_q}.
 \label{eq:leakage_input_referred}
\end{equation}
Here $h$ is Planck's constant and $\nu_q$ is the quantum optical-carrier
frequency. Powers and frequencies are converted to SI units when
Eq.~\eqref{eq:leakage_input_referred} is evaluated numerically.
The common demultiplexer and fiber transmissions cancel when the leakage power is referred to the channel input. The factor of two converts the admitted mean photon number in the selected polarization into input-referred quadrature excess noise in the adopted SNU convention \cite{Kumar2015,Laudenbach2018}. Polarization extinction can depend on frequency \cite{Penninckx2005}; the numerical model adopts a constant $\mathrm{PER}=30~\mathrm{dB}$ over the modeled bands. The residual QPSK waveform is taken to be zero mean and statistically
independent of the quantum symbols. Its admitted second moment is
represented as additive Gaussian noise in the channel model used
for the mutual-information and covariance calculations.

The nominal $S_c$ describes the intended pulse-shaped waveform. To test additional transmitter noise outside the intended classical signal band (out-of-band, OOB), let $S_{\rm OOB}(f)$ denote its optical power spectral density at the channel input, independent of the
quantum symbols. Its fraction admitted by the quantum filter before polarization suppression is
\begin{equation}
 \epsilon_{\rm OOB}
 =\frac{1}{P_c}\int S_{\rm OOB}(f)|H_q(f)|^2\,df.
 \label{eq:oob_fraction}
\end{equation}
The sensitivity calculation replaces $\Omega$ in Eq.~\eqref{eq:leakage_input_referred} by $\Omega+\epsilon_{\rm OOB}$, with the same polarization suppression. The nominal case uses $\epsilon_{\rm OOB}=0$; finite values represent additional sharing-induced noise above the calibrated baseline. The dimensionless fraction $\epsilon_{\rm OOB}$ represents transmitter-noise power integrated through the quantum receive filter and normalized by the classical launch power, and can be obtained from a measured transmitter noise spectrum. Its effect is examined in Section~\ref{subsec:noise_margin}.

For a forward-propagating classical channel, the input-referred spontaneous Raman excess noise reduces to \cite{Kumar2015}
\begin{equation}
 \xi_{\rm Raman}
 =\frac{\lambda_q^3}{hc_0^2}\beta_RP_cL.
 \label{eq:raman_compact}
\end{equation}
Here $\lambda_q$ is the quantum wavelength, $c_0$ is the speed of light, and
$\beta_R$ is the polarization-summed Raman scattering coefficient in the convention of \cite{Kumar2015}. Selection of one polarization contributes a factor $1/2$, which cancels the factor $2$ in the photon-to-SNU conversion and yields Eq.~\eqref{eq:raman_compact}. When $\beta_R$ is tabulated
in $\mathrm{km^{-1}nm^{-1}}$, the SI quantities entering
Eq.~\eqref{eq:raman_compact} are
\begin{align}
 \widetilde\beta_R&=10^6\beta_R\ \mathrm{m^{-2}},\notag\\
 P_c[\mathrm W]&=10^{-3}P_c[\mathrm{mW}],\notag\\
 L[\mathrm m]&=10^3L[\mathrm{km}].
 \label{eq:raman_si_conversion}
\end{align}
Section~\ref{subsec:reference_configuration} evaluates
Eq.~\eqref{eq:raman_compact} with these conversions, while the channel
transmittance entering the security covariance is set independently by the
platform calibration.

These input-referred sharing contributions must next be combined with the
electronic noise introduced by the trusted receiver.

\subsection{Bandwidth-dependent trusted receiver noise}
\label{subsec:receiver_noise_bandwidth_model}

We next separate the shot-noise reference from the trusted electronic noise
and model how the latter changes with receiver bandwidth. Receiver
characterization integrates the electronic-
and shot-noise power spectral densities (PSDs) over the detection bandwidth
\cite{RicardThesis2025,RicardNoiseCalibration2025}. Let $S_{\rm el}(f)$ and
$S_{\rm SN}(f)$ denote these PSDs in the same electrical-variance units per
unit frequency. We describe their bandwidth dependence using a two-term electronic-noise model and a white shot-noise spectrum,
\begin{equation}
 S_{\rm el}(f)=S_{{\rm el},0}+S_{{\rm el},2}f^2,
 \qquad
 S_{\rm SN}(f)=S_{{\rm SN},0},
 \label{eq:receiver_noise_psds}
\end{equation}
where $S_{{\rm el},0}$ is the white electronic-noise floor,
$S_{{\rm el},2}$ the quadratic electronic-noise coefficient, and
$S_{{\rm SN},0}$ the white shot-noise level. Frequency is expressed in GHz in
this fitted model. For the symmetric modeled
passband $[-B_{\rm RX}/2,B_{\rm RX}/2]$, the band-integrated electronic- and
shot-noise variances are
\begin{align}
 V_{\rm el}(B_{\rm RX})
 &=S_{{\rm el},0}B_{\rm RX}+\frac{S_{{\rm el},2}}{12}B_{\rm RX}^3,
 \notag\\
 V_{\rm SN}(B_{\rm RX})
 &=S_{{\rm SN},0}B_{\rm RX}.
 \label{eq:receiver_noise_integrals}
\end{align}
Normalizing the integrated electronic variance by the shot-noise variance gives
\begin{equation}
 \xi_{\rm rec}(B_{\rm RX})
 =\frac{V_{\rm el}(B_{\rm RX})}{V_{\rm SN}(B_{\rm RX})}
 =\xi_0+\kappa_{\rm RX}B_{\rm RX}^2.
 \label{eq:receiver_noise_bandwidth_law}
\end{equation}
Here $V_{\rm el}$ and $V_{\rm SN}$ are expressed in the same electrical-variance units and $B_{\rm RX}$ is expressed in GHz.
The coefficients $\xi_0=0.07330~\mathrm{SNU}$ and $\kappa_{\rm RX}=0.13636~\mathrm{SNU/GHz^2}$ are determined by the adopted points $(0.35~\mathrm{GHz},0.09~\mathrm{SNU})$ and $(0.75~\mathrm{GHz},0.15~\mathrm{SNU})$. The first combines the baseline bandwidth budget with the receiver-noise parameter in \cite{RicardThesis2025}. The second design point uses the $0.15$-SNU electronic-noise value reported in \cite{PanHighRate2022} as a reference for an illustrative receiver scenario. Pan et al. report a balanced-detector bandwidth of $1.6~\mathrm{GHz}$ and a $1.3~\mathrm{GHz}$ quantum-signal filter; their $750~\mathrm{MHz}$ value is a digital frequency shift of the reference signal. In our model, $B_{\rm RX}=0.75~\mathrm{GHz}$ is the assumed receiver-bandwidth budget for this illustrative scenario.

The input-referred link excess noise is
\begin{equation}
 \xi_{\rm link}=\xi_A+\xi_{\rm leak}+\xi_{\rm Raman}.
 \label{eq:link_excess_noise}
\end{equation}
The detected-noise budget contains separate shot-noise, trusted electronic-noise, and link excess-noise contributions, with the shot-noise variance normalized to $N_0=1~\mathrm{SNU}$. After
reference-plane conversion, the variance of the additive noise in each
detected heterodyne quadrature is
\begin{equation}
 \begin{aligned}
 \sigma_N^2(B_{\rm RX})
 &=N_0+\xi_{\rm rec}(B_{\rm RX})
 +\frac{\eta_BT_{\rm ch}}{2}\xi_{\rm link},\\
 N_0&=1~\mathrm{SNU}.
 \end{aligned}
 \label{eq:detected_noise_budget}
\end{equation}
Thus $\xi_{\rm rec}$ is trusted receiver noise, whereas $\xi_{\rm link}$ is
the channel-input-referred contribution used in the covariance that bounds
Eve's information. Appendix~\ref{app:security_backend} gives the corresponding
trusted-receiver purification. With the link, shot-noise, and electronic-noise
terms separated, the discrete-modulation key rate can now be evaluated.

\subsection{Discrete-modulation key rate with a trusted receiver}
\label{subsec:quantum_rate_model}

The secret-key rate is obtained from the binomial square 64-QAM distribution, for which the analytical covariance lower bound can be evaluated explicitly \cite{Denys2021}. For $k,\ell\in\{0,\ldots,7\}$,
\begin{equation}
 p_{k\ell}=\frac{\binom{7}{k}\binom{7}{\ell}}{2^{14}},
 \qquad
 \alpha_{k\ell}=\sqrt{\frac{V_A}{7}}
 \left[k-\frac{7}{2}+i\!\left(\ell-\frac{7}{2}\right)\right].
 \label{eq:binomial_64qam}
\end{equation}
We model the platform's 64-point constellation using the binomial probabilities in Eq.~\eqref{eq:binomial_64qam}. This normalization gives $2\sum_{k,\ell}p_{k\ell}|\alpha_{k\ell}|^2=V_A$. Appendix~\ref{app:security_backend} gives the covariance normalization, trusted-receiver purification, and symplectic-eigenvalue evaluation of $\chi(E;Y)$.

Let $X=(X_I,X_Q)$ denote Alice's transmitted 64-QAM symbol,
$Y=(Y_I,Y_Q)$ Bob's detected heterodyne outcome, and
$N=(N_I,N_Q)$ the additive noise after reference-plane conversion. For
$u\in\{I,Q\}$, the equivalent real-valued channel for each measured quadrature is
\begin{equation}
 Y_u=\sqrt{\frac{\eta_BT_{\rm ch}}{2}}X_u+N_u.
 \label{eq:discrete_detected_channel}
\end{equation}
For the real eight-level pulse-amplitude modulation (8-PAM) component, let
\(p_j=\binom{7}{j}/2^7\) and
\(x_j=2\sqrt{V_A/7}\,(j-7/2)\) denote, respectively, the probability and
quadrature displacement of symbol \(j\) in the adopted SNU convention. Conditional on transmitting $x_j$, the measured value is Gaussian,
\begin{equation}
 \begin{aligned}
 p(y|x_j)&=\frac{1}{\sqrt{2\pi\sigma_N^2}}
 \exp\!\left[-\frac{(y-g_{\rm ch}x_j)^2}{2\sigma_N^2}\right],\\
 g_{\rm ch}&=\sqrt{\frac{\eta_BT_{\rm ch}}{2}}.
 \end{aligned}
 \label{eq:conditional_output_density}
\end{equation}
where $\sigma_N^2=\mathrm{Var}(N_u)$ is given by
Eq.~\eqref{eq:detected_noise_budget}. Thus $p(y|x_j)$ is the probability density
of Bob observing $y$ when Alice transmits $x_j$, and the unconditional output
density is the mixture $p(y)=\sum_kp_kp(y|x_k)$.
Let $\mathcal B_m$, $m\in\{0,\ldots,63\}$, be the scalar quantizer cells
defined below Eq.~\eqref{eq:joint_quantized_heterodyne_outcome}. Their exact
conditional and marginal probabilities are
\begin{equation}
 p_{m|j}=\int_{\mathcal B_m}p(y|x_j)\,dy,
 \qquad
 p_m=\sum_{j=0}^{7}p_jp_{m|j}.
 \label{eq:quantized_cell_probabilities}
\end{equation}
Accordingly, the two joint quantities used in
Eq.~\eqref{eq:reconciliation_disclosure} are evaluated as
\begin{align}
 H(\mathbf X_B)&=-2\sum_{m=0}^{63}p_m\log_2p_m,
 \label{eq:joint_quantized_entropy}\\
 I(X;\mathbf X_B)&=2\sum_{j=0}^{7}\sum_{m=0}^{63}
 p_jp_{m|j}\log_2\!\frac{p_{m|j}}{p_m}.
 \label{eq:joint_quantized_mutual_information}
\end{align}
The Gaussian cumulative distribution is used to integrate each cell,
including the two saturated tails. Since square 64-QAM factorizes into two
independent, identical 8-PAM quadratures, the total continuous-output
discrete-input mutual information used in the security bound is
\begin{equation}
 I(X;Y)
 =2\sum_{j=0}^{7}p_j\int p(y|x_j)
 \log_2\!\frac{p(y|x_j)}{\sum_kp_kp(y|x_k)}\,dy,
 \label{eq:discrete_mutual_information}
\end{equation}
which is evaluated using 32-point Gauss--Hermite quadrature. The interpolation and convergence checks used for the repeated simulation calls are given in Appendix~\ref{app:numerical_validation}. The asymptotic reverse-reconciliation secret fraction and throughput are \cite{Denys2021}
\begin{align}
 r_q&=\left[\beta I(X;Y)-\chi(E;Y)\right]_+,
 \label{eq:secret_fraction}\\
 K_q&=10^3f_q(1-\nu_{\rm PE})r_q\quad\mathrm{Mbit/s}.
 \label{eq:secret_key_throughput}
\end{align}
The resulting $K_q$ enters the service constraints in Eq.~\eqref{eq:scr_service_constraints}.

\subsection{Nested optimization and resource criterion}
\label{subsec:scr_optimization_problem}

Combining the preceding rates and physical constraints gives the nested optimization over the shared link. The vector of optimization variables is

\begin{equation}
 \mathbf x=(L,f_c,f_q,\Delta f,V_A,P_c,s).
 \label{eq:optimizer_vector}
\end{equation}
For each fixed $(L,P_c)$, the inner problem maximizes $s$ subject to Eqs.~\eqref{eq:scr_service_constraints}, \eqref{eq:inner_spectral_constraint}, and \eqref{eq:outer_receiver_constraints}, together with
\begin{equation}
 \mathrm{SNR}_{c,\mathrm{dB}}\geq14\ \mathrm{dB},
 \qquad r_q\geq0.
 \label{eq:transport_security_constraints}
\end{equation}
The distance-specific and global optima are
\begin{align}
 \mathrm{SCR}^\star(L)
 &=\max_{P_c,f_c,f_q,\Delta f,V_A,s}s,
 \label{eq:distance_scr_optimization}\\
 \mathrm{SCR}_{\rm glob}^\star
 &=\max_{25\leq L\leq100\,\mathrm{km}}\mathrm{SCR}^\star(L).
 \label{eq:nested_scr_optimization}
\end{align}
For the receiver study at fixed distance, the corresponding conditioned
shared optimum is
\begin{align}
 \mathrm{SCR}_{\rm sh}^{\star}(B_{\rm RX};L)
 &=\max_{\mathbf{x}_{\rm RX}}
 \mathrm{SCR}_{\rm sh}\!\left(\mathbf{x}_{\rm RX};\right.\notag\\
 &\hspace{4.4em}\left.B_{\rm RX},\xi_{\rm rec}(B_{\rm RX}),L\right),
 \label{eq:receiver_conditioned_scr}
\end{align}
where $\mathbf{x}_{\rm RX}=(f_c,f_q,\Delta f,V_A,P_c)$. All constraints of
Section~\ref{sec:optimizer_model}, including
$f_q\leq B_{\rm RX}/(1+\rho_q)$, are retained.

At fixed $(f_c,L)$, the classical throughput in Eq.~\eqref{eq:classical_net_rate} is constant after Eq.~\eqref{eq:transport_security_constraints} is met, whereas Eqs.~\eqref{eq:leakage_input_referred} and \eqref{eq:raman_compact} are nondecreasing in $P_c$. Consequently, the joint distance optimizer sets $P_c$ at the minimum value satisfying the SNR constraint and the declared launch-power lower bound for each candidate $f_c$. A compact-spectrum selection rule chooses the smallest $\Delta f$ that retains at least $99.9\%$ of the maximum SCR; the resulting rate deviation is bounded by $0.1\%$ by construction.

The separated reference is provisioned with sufficient classical capacity
to carry the support traffic and the OTP-encrypted payload. Its secure
payload is therefore key-limited and equals
\(\mathrm{SKR}_{\rm sep}^{\star}\). For the equal-weight index of the modeled transmission direction, the shared architecture uses four resources and the separated reference uses six. The shared architecture satisfies the shared-versus-separated rate-per-resource condition when
\begin{align}
 \frac{\mathrm{SCR}_{\rm sh}^\star}{4}
 &\geq\frac{\mathrm{SKR}_{\rm sep}^\star}{6}
 \notag\\
 \Longleftrightarrow\qquad
 \mathcal{P}_{\rm sh}
 &=10\log_{10}\!\left(
 \frac{\mathrm{SKR}_{\rm sep}^\star}{\mathrm{SCR}_{\rm sh}^\star}
 \right)
 \notag\\
 &\leq10\log_{10}\!\left(\frac{6}{4}\right)
 =1.761~\mathrm{dB}.
 \label{eq:shared_separated_hardware_threshold}
\end{align}
The $1.761$-dB threshold in Eq.~\eqref{eq:shared_separated_hardware_threshold} follows from the equal-weight resource index adopted here. For symmetric two-way architectures, the corresponding counts become $2\times4=8$ for shared transmission and $2\times6=12$ for separated transmission, preserving the same resource ratio when the compared rates satisfy the coupled two-way transport constraints in Eq.~\eqref{eq:two_way_transport_constraints}. If instead the same additional return-resource count $C_0$ is included in both one-direction architectures, the corresponding threshold is
\begin{equation}
 \mathcal P_{\rm sh}\leq10\log_{10}\!\left(\frac{6+C_0}{4+C_0}\right).
 \label{eq:full_resource_threshold}
\end{equation}
For example, adding one independent transmitter, one fiber, and one receiver/DSP chain to each gives $C_0=3$ and a threshold of $1.091~\mathrm{dB}$.

\section{Hardware, Spectral, Receiver, and Power Co-Design Study}
\label{sec:codesign_study}

The co-design study identifies which physical constraint limits the secure
payload in each operating regime. It begins with the documented
platform inputs and numerical protocol and then examines spectral separation,
distance and launch power, receiver bandwidth, and shared-noise margin. All
sweeps use the same numerical implementation and can be reproduced with the
accompanying numerical package.

\subsection{Platform parameters and numerical protocol}
\label{subsec:reference_configuration}

Selected measurements and calibrations from the platform characterization are used as simulation inputs \cite{RicardThesis2025}. Table~\ref{tab:platform_parameters} lists the reference receiver configuration and the remaining design quantities. The characterization reports an average excess-noise value of $4.3~\mathrm{mSNU}$ together with $T_{\rm ch}=0.28$ at $25~\mathrm{km}$. In the reference-plane convention adopted here, this value is denoted by $\xi_B$ and treated as a channel-output-referred baseline excess-noise contribution before trusted detection. The input-referred quantity used in the channel covariance is therefore $\xi_A=\xi_B/T_{\rm ch}$. After trusted detection, it contributes $\eta_BT_{\rm ch}\xi_A/2$ to each measured heterodyne quadrature, consistently with Appendix~\ref{app:security_backend}. The distance law uses the reported 25-km transmittance as its reference point,
\begin{equation}
 T_{\rm ch}=0.28\times10^{-0.2(L-25)/10},
 \label{eq:distance_transmittance}
\end{equation}
where $0.2~\mathrm{dB/km}$ is the distributed fiber-loss slope inferred from the reported $5$-dB fiber attenuation over $25~\mathrm{km}$. The calibrated $T_{\rm ch}=0.28$ corresponds to $5.528$ dB total loss. Equivalently, Eq.~\eqref{eq:distance_transmittance} is $T_{\rm ch}=0.88544\times10^{-0.2L/10}$, retaining the residual fixed calibration factor separately from the distance-dependent attenuation. Here $\eta_{\rm rec}$ is the optical receiver transmission, $\eta_{\rm det}$ the detector efficiency, and $\eta_{\rm DSP}$ the DSP-mode efficiency. Their product gives the trusted efficiency,
\begin{equation}
 \eta_B=\eta_{\rm rec}\eta_{\rm det}\eta_{\rm DSP}
 =0.60\times0.88\times0.884=0.466752.
 \label{eq:trusted_efficiency_value}
\end{equation}

\begin{table}[!t]
\caption{Platform and numerical parameters.}
\label{tab:platform_parameters}
\centering
\scriptsize
\setlength{\tabcolsep}{2.5pt}
\begin{tabularx}{\columnwidth}{@{}ll>{\raggedright\arraybackslash}X@{}}
\toprule
Quantity & Value & Basis \\
\midrule
$T_{\rm ch}$ at $L=25~\mathrm{km}$ & $0.28$ & platform \\
$\xi_B$ at $L=25~\mathrm{km}$ & $4.3$ mSNU & platform \\
$\xi_A=\xi_B/T_{\rm ch}$ & $15.36$ mSNU & derived \\
$\xi_{\rm rec}(0.35~\mathrm{GHz})$ & $0.09$ SNU & Ref.~\cite{RicardThesis2025} \\
$\eta_{\rm rec},\eta_{\rm det},\eta_{\rm DSP}$ & $0.60,0.88,0.884$ & platform \\
$B_{\rm RX}$ & $0.35$ GHz & adopted budget \\
$\rho_c,\rho_q$ & $0.10,0.30$ & platform \\
$\beta,\nu_{\rm PE}$ & $0.95,0.10$ & protocol assumptions \\
$\beta_R$ & $3.0\times10^{-9}~\mathrm{km^{-1}nm^{-1}}$ & representative design value \cite{Kumar2015} \\
$\lambda_q$ & $1550$ nm & platform \\
PER, filter order & $30$ dB, $4$ & design \\
$B_{\rm tot},B_{\rm guard}$ & $6.8,0.12$ GHz & design \\
$\eta_c$, $R_{\rm back}$ & $0.95$, $10$ Gbit/s & design; independent net return capacity \\
$f_{\rm EC}$, $\epsilon_{\rm OOB}$ & $1/0.95$, $0$ & nominal design \\
$P_{c,0}$ & $1$ mW & assumed power reference \\
\bottomrule
\end{tabularx}
\end{table}

To evaluate the spectral-overlap integral in
Eq.~\eqref{eq:spectral_overlap}, the quantum-channel receive filter is represented by
an $n_f$th-order Butterworth low-pass power response \cite{Butterworth1930},
\begin{equation}
 |H_q(f)|^2
 =\left[
 1+\left(\frac{|f|}{f_{\rm cut}}\right)^{2n_f}
 \right]^{-1},
 \qquad
 f_{\rm cut}=\frac{B_q}{2}.
 \label{eq:numerical_filter}
\end{equation}
We use this Butterworth response as an effective receive-filter model for the spectral-overlap calculation. Its cutoff is placed
at half the occupied quantum bandwidth, and the nominal filter
order is $n_f=4$. The response is a design assumption rather than
a measured transfer function of the complete receiver/DSP chain. The classical-signal PSD
$S_c(f)$ is the unit-area raised-cosine PSD defined by the
classical symbol rate $f_c$ and roll-off $\rho_c$. Together, $S_c(f)$ and
$H_q(f)$ determine the admitted classical-power fraction
$\Omega(\Delta f)$ through Eq.~\eqref{eq:spectral_overlap}.

The leakage contribution also depends on the polarization suppression
$\eta_{\rm pol}=10^{-\mathrm{PER}/10}$. The nominal design uses
$\mathrm{PER}=30~\mathrm{dB}$ and a fourth-order receive filter. These
quantities enter the physical leakage model in
Eqs.~\eqref{eq:spectral_overlap}--\eqref{eq:leakage_input_referred}.

The second shared-link contribution is forward spontaneous Raman
scattering. The numerical implementation evaluates
Eq.~\eqref{eq:raman_compact} directly, with the unit conversions in
Eq.~\eqref{eq:raman_si_conversion}. For the wavelength in
Table~\ref{tab:platform_parameters}, the implemented expression is
\begin{equation}
\begin{aligned}
 \xi_{\rm Raman}
 &=\frac{(1.550\times10^{-6})^3}{
 (6.62607015\times10^{-34})(299792458)^2}\\
 &\quad\times(10^6\beta_R)(10^{-3}P_c)(10^3L)\\
 &=4.68985\times10^{-3}
 \left(
 \frac{\beta_R}
 {3.0\times10^{-9}\,\mathrm{km^{-1}nm^{-1}}}
 \right )\\
 &\quad\times
 \left(\frac{P_c}{1\,\mathrm{mW}}\right)
 \left(\frac{L}{25\,\mathrm{km}}\right).
\end{aligned}
 \label{eq:raman_numerical_normalization}
\end{equation}
Here $P_c$ and $L$ in the second line are the numerical values in milliwatts
and kilometers, respectively. Thus the analytical model gives
$4.68985~\mathrm{mSNU}$ for $P_c=1~\mathrm{mW}$, $L=25~\mathrm{km}$, and
$\beta_R=3.0\times10^{-9}~\mathrm{km^{-1}nm^{-1}}$. The main simulations
use this coefficient, while the reported
$(1.5$--$3.1)\times10^{-9}~\mathrm{km^{-1}nm^{-1}}$ interval defines a
Raman-coefficient sensitivity scenario. The coefficient range is taken from measurements on wavelength-separated WDM channels \cite{Kumar2015}. Its use at GHz-scale detunings is an assumed extrapolation examined through the coefficient sensitivity sweep.

The classical feasibility condition is $\mathrm{SNR}_{c,\mathrm{dB}}\geq14~\mathrm{dB}$, following the platform region with zero observed errors over $2\times10^6$ QPSK symbols \cite{RicardThesis2025}. At the adopted symbol SNR $E_s/N_0=14$ dB, ideal uncoded QPSK has a bit-error probability of approximately $2.7\times10^{-7}$, so reliable transport requires coding or error detection and retransmission within the assumed efficiency budget in Eq.~\eqref{eq:classical_net_rate}. The classical-power model uses the SNR--relative-power data in Fig.~4.10 of \cite{RicardThesis2025}. A least-squares fit to $15$ digitized points gives
\begin{equation}
 P_{c,\mathrm{rel,dB}}
 =1.025073\,\mathrm{SNR}_{c,\mathrm{dB}}-20.208927.
 \label{eq:classical_snr_fit}
\end{equation}
Here $P_{c,\mathrm{rel,dB}}=10\log_{10}(P_c/P_{c,0})$ denotes the relative classical launch power in decibels with a root-mean-square error (RMSE) of $0.262~\mathrm{dB}$. We adopt the absolute-power reference $P_{c,0}=1~\mathrm{mW}$ as a simulation assumption and use the fit at $14~\mathrm{dB}$
to define
\begin{equation}
 P_{c,14}
 =P_{c,0}\,10^{(14\times1.025073-20.208927)/10}.
 \label{eq:classical_reference_power}
\end{equation}
The reported absolute launch powers and the resulting leakage
and Raman contributions are conditional on this assumed reference.
For other distances and classical symbol rates, we use
$\mathrm{SNR}_c\propto P_cT_{\rm ch}/f_c$
under the adopted receiver-noise assumptions \cite{Agrawal2002}:
\begin{align}
 \mathrm{SNR}_{c,\mathrm{dB}}
 &=14+10\log_{10}\!\left(\frac{P_c}{P_{c,14}}\right)
 \notag\\
 &\quad+10\log_{10}\!\left(\frac{T_{\rm ch}}{0.28}\right)
 +10\log_{10}\!\left(\frac{4~\mathrm{GBd}}{f_c}\right).
 \label{eq:classical_snr_model}
\end{align}

The bounded baseline search uses $0.05\leq f_c\leq4~\mathrm{GBd}$, $0.05\leq f_q\leq0.26923~\mathrm{GBd}$, $0.35\leq\Delta f\leq6~\mathrm{GHz}$, $0.5\leq V_A\leq12~\mathrm{SNU}$, $0.004\leq P_c\leq4~\mathrm{mW}$, and $25\leq L\leq100~\mathrm{km}$. The $f_c$ ceiling is the platform rate, and the $f_q$ ceiling follows Eq.~\eqref{eq:outer_receiver_constraints} at $B_{\rm RX}=0.35~\mathrm{GHz}$; the receiver co-design study updates this upper bound for each selected bandwidth. The remaining endpoints provide numerical coverage of the feasible and noise-limited regions. Fock-space truncation, mutual-information integration, physicality of the covariance matrices, and the ideal-receiver limit are verified in Appendix~\ref{app:numerical_validation}.

At $25~\mathrm{km}$, the separated-link optimization gives
\begin{equation}
 V_A^\star=4.762~\mathrm{SNU},
 \qquad
 \mathrm{SKR}_{\rm sep}^\star=12.095~\mathrm{Mbit/s}.
 \label{eq:separated_reference_result}
\end{equation}
The reference symplectic spectrum and convergence checks are reported in Appendix~\ref{app:numerical_validation}.

\subsection{Spectral design at 25 km}

At $25~\mathrm{km}$, the spectral separation determines how strongly the classical waveform couples into the quantum-channel receive filter. We therefore evaluate the optimized SCR as $\Delta f$ varies, reoptimizing $(f_c,f_q,V_A,P_c)$ at each separation. The receiver constraint is active throughout the high-rate region,
\begin{equation}
 f_q^\star=\frac{0.35}{1+0.30}=0.26923~\mathrm{GBd}.
 \label{eq:receiver_limited_symbol_rate}
\end{equation}

Near the spectral-packing boundary, increasing channel separation rapidly reduces classical spectral leakage, whereas at larger separations the Raman contribution and receiver-bandwidth constraint limit further SCR improvement.

Applying the compact-spectrum rule to the maximum SCR of $12.091~\mathrm{Mbit/s}$ selects $\Delta f^\star=1.197~\mathrm{GHz}$, retaining at least $99.9\%$ of the maximum with an SCR of $12.079~\mathrm{Mbit/s}$. The associated leakage and Raman components are $0.073$ and $0.028~\mathrm{mSNU}$, respectively. At this point, the joint quantizer gives $H(\mathbf X_B)=10.0956$ and $I(X;\mathbf X_B)=0.3615~\mathrm{bit/symbol}$, hence $\lambda_{\rm EC}=10.2465~\mathrm{bit/symbol}$.

The support traffic is $R_{\rm sup}=2644.340~\mathrm{Mbit/s}$, approximately $219$ times the OTP payload rate. It consists of $R_{\rm PE}=161.538~\mathrm{Mbit/s}$ from Alice to Bob and $R_{\rm EC}=2482.802~\mathrm{Mbit/s}$ from Bob to Alice on the opposite fiber.

At this optimized point, parameter-estimation disclosure and the OTP payload require $161.538+12.079=173.617~\mathrm{Mbit/s}$ of forward throughput, corresponding to $f_c\gtrsim0.09138~\mathrm{GBd}$ at $\eta_c=0.95$. If both directions were operated with the same parameters at the same time, each classical channel would also have to carry about $2.483~\mathrm{Gbit/s}$ of reconciliation traffic for the opposite quantum transmission. The resulting load is approximately $2.656~\mathrm{Gbit/s}$ per direction, requiring $f_c\gtrsim1.398~\mathrm{GBd}$. These additional transport requirements define a two-way operating point distinct from the one-direction optimum in Table~\ref{tab:distance_points}, requiring reoptimization using the coupled constraints in Eq.~\eqref{eq:two_way_transport_constraints}.

\subsection{Joint distance and classical-power optimization}

After establishing the spectral behavior at $25~\mathrm{km}$, we extend the optimization to examine how the required classical launch power changes with distance. The joint sweep evaluates the optimum at each distance over $25$--$100~\mathrm{km}$. At each distance, the optimizer selects the minimum power satisfying the $14$-dB classical feasibility constraint; further launch power would increase the Raman contribution without increasing the available classical throughput.

Table~\ref{tab:distance_points} lists the corresponding optimization variables, separated-link reference rates, shared-link rates, and penalties evaluated with Eq.~\eqref{eq:shared_separated_hardware_threshold}.

\begin{table*}[!t]
\caption{Jointly optimized one-direction distance results with independently provisioned return capacity. For each fixed distance, the quantum and classical optimization variables are first optimized for maximum SCR; the smallest spectral separation retaining at least $99.9\%$ of that maximum is then selected. The penalty is calculated from the full-precision separated and shared rates. The absolute launch powers follow the assumed $P_{c,0}=1~\mathrm{mW}$ reference described in
Section~\ref{subsec:reference_configuration}. \textsuperscript{*} Values are optimized.}
\label{tab:distance_points}
\centering
\scriptsize
\setlength{\tabcolsep}{3.5pt}
\renewcommand{\arraystretch}{1.12}
\begin{tabular}{@{}r r r r r r r r r p{2.4cm}@{}}
\toprule
$L$ & $P_c^{*}$ & $f_c^{*}$ & $f_q^{*}$ & $\Delta f^{*}$ & $V_A^{*}$ &
$\mathrm{SKR}_{\rm sep}^{*}$ & $\mathrm{SCR}_{\rm sh}^{*}$ & Penalty & $(\xi_{\rm leak},\xi_{\rm Raman})$ \\
(km) & (mW) & (GBd) & (GBd) & (GHz) & (SNU) & (Mbit/s) & (Mbit/s) & (dB) & (mSNU) \\
\midrule
25 & 0.0059 & 0.091 & 0.269 & 1.197 & 4.762 & 12.095 & 12.079 & 0.006 & $(0.073,0.028)$ \\
50 & 0.0178 & 0.087 & 0.269 & 1.423 & 3.953 & 2.866 & 2.855 & 0.018 & $(0.054,0.167)$ \\
75 & 0.0554 & 0.085 & 0.269 & 1.695 & 3.674 & 0.791 & 0.776 & 0.085 & $(0.042,0.780)$ \\
100 & 0.1747 & 0.085 & 0.269 & 2.028 & 3.571 & 0.227 & 0.205 & 0.444 & $(0.031,3.277)$ \\
\bottomrule
\end{tabular}
\end{table*}

The outer optimization selects $L^\star=25~\mathrm{km}$, the lower boundary of the evaluated interval. Maintaining the Alice-to-Bob classical SNR while carrying parameter-estimation data and the payload as the distance increases from $25$ to $100~\mathrm{km}$ raises the required launch power from $0.0059$ to $0.1747~\mathrm{mW}$, with the corresponding Raman contribution increasing from $0.028$ to $3.277~\mathrm{mSNU}$. At $100~\mathrm{km}$, the optimized quantum symbol rate is $0.269~\mathrm{GBd}$ and the sharing penalty is $0.444~\mathrm{dB}$.

\begin{figure*}[!t]
\centering
\includegraphics[width=0.94\textwidth]{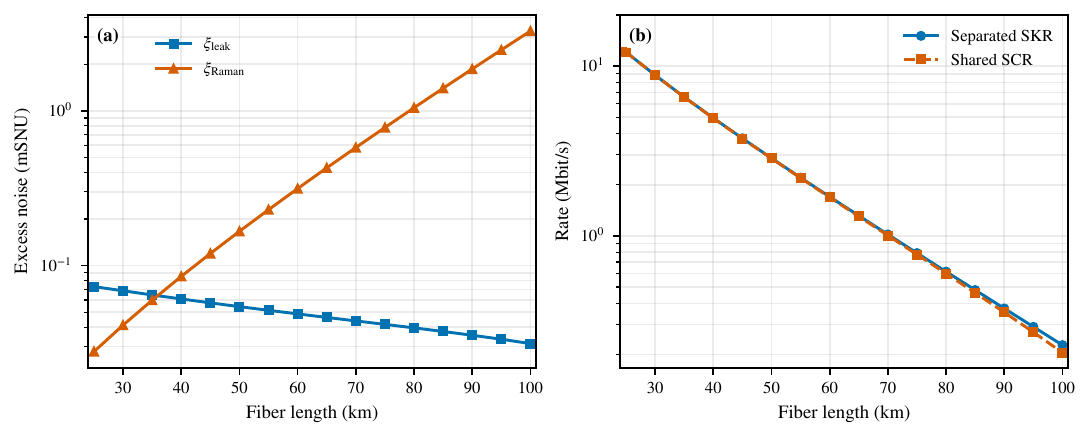}
\caption{\parbox{0.90\textwidth}{\raggedright
Distance dependence of the optimized one-direction link under provisioned return capacity. (a) Leakage and forward Raman excess noise versus fiber length. (b) Separated-link SKR and shared-link SCR versus fiber length. The shared-link optimization variables are reoptimized at each distance.}}
\label{fig:shared_noise_components}
\end{figure*}

Fig.~\ref{fig:shared_noise_components} brings these two trends together. Panel~(a) shows the leakage and Raman contributions created by hardware sharing, while panel~(b) compares the separated-link SKR and shared-link SCR over the full $25$--$100~\mathrm{km}$ interval.

\subsection{Receiver-bandwidth scaling at 25 km}
\label{subsec:receiver_bandwidth_results}

The distance sweep above uses the adopted budget
$B_{\rm RX}=0.35~\mathrm{GHz}$. To examine the bandwidth--noise
trade-off, we reoptimize the $25$-km configuration at
$B_{\rm RX}\in\{0.35,0.50,0.75\}~\mathrm{GHz}$ using
Eq.~\eqref{eq:receiver_conditioned_scr}.
The first value is the baseline modeling point, while the latter
two define illustrative receiver-bandwidth scenarios.
At each point, the electronic noise follows the assumed relation
in Eq.~\eqref{eq:receiver_noise_bandwidth_law};
$(f_c,f_q,\Delta f,V_A,P_c)$ is reoptimized; and $\eta_B$ and
$\xi_A$ remain fixed. Optimizing the separated reference with the
same receiver bandwidth and electronic noise isolates the rate
penalty associated with transmitter and fiber sharing.

\begin{figure*}[!t]
\centering
\includegraphics[width=0.94\textwidth]
{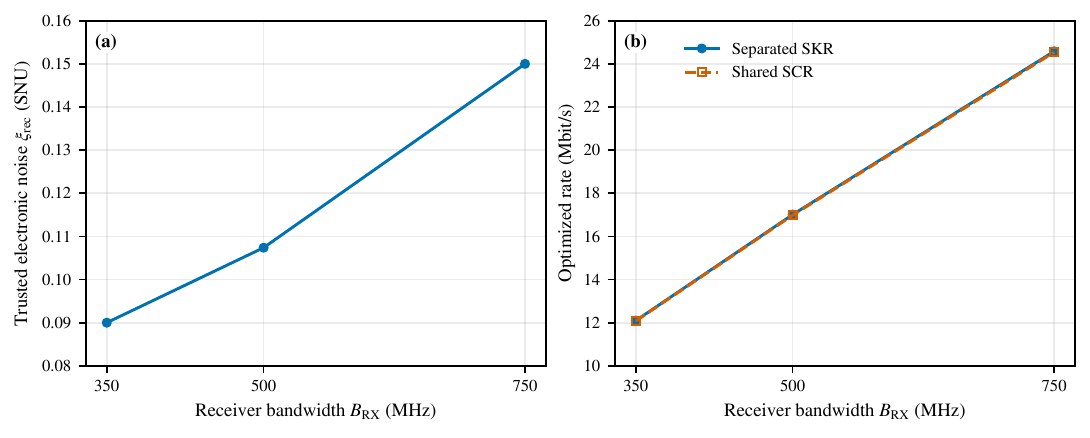}
\caption{\parbox{0.92\textwidth}{\raggedright
Receiver-bandwidth and electronic-noise co-design at $L=25~\mathrm{km}$. (a) Trusted electronic noise obtained from Eq.~\eqref{eq:receiver_noise_bandwidth_law}. (b) Optimized separated-link SKR and shared-link SCR at the three selected receiver bandwidths. The $350$-MHz point combines the adopted baseline bandwidth budget with the $0.09$-SNU receiver-noise parameter; the $500$- and $750$-MHz points are illustrative bandwidth--noise scenarios.}}
\label{fig:receiver_bandwidth_codesign}
\end{figure*}

Figure~\ref{fig:receiver_bandwidth_codesign}(a) shows that the assumed trusted electronic noise increases from $0.090~\mathrm{SNU}$ at $350~\mathrm{MHz}$ to $0.107~\mathrm{SNU}$ at $500~\mathrm{MHz}$ and $0.150~\mathrm{SNU}$ at $750~\mathrm{MHz}$. The receiver constraint remains active, giving $f_q^\star=0.269$, $0.385$, and $0.577~\mathrm{GBd}$, respectively. The corresponding shared SCR values are $12.079$, $16.984$, and $24.535~\mathrm{Mbit/s}$, as shown in Fig.~\ref{fig:receiver_bandwidth_codesign}(b). For these assumed bandwidth--noise scenarios, the additional symbol-rate budget outweighs the loss from increased trusted noise.

The sharing penalties under matched receiver assumptions are $0.0060$, $0.0067$, and $0.0079~\mathrm{dB}$. The maximum Bob-to-Alice reconciliation traffic is $5.330~\mathrm{Gbit/s}$, within the independently provisioned $10$-Gbit/s net return service.

\subsection{Noise margin and hardware interpretation}
\label{subsec:noise_margin}

At the $350$-MHz baseline, we evaluate the additional sharing-induced excess noise compatible with the hardware-normalized rate criterion and with positive key generation. At $25~\mathrm{km}$, a sensitivity sweep over this excess noise gives the thresholds in Table~\ref{tab:noise_boundaries}. Each entry reoptimizes $V_A$ with the trusted-receiver key-rate calculation. Here, \(\xi_{\rm sh}\) denotes an additional input-referred excess-noise
contribution associated with hardware sharing, above the platform
baseline. For the physical shared link considered here,
\(\xi_{\rm sh}=\xi_{\rm leak}+\xi_{\rm Raman}\).

\begin{table}[!t]
\caption{Sharing-induced excess-noise thresholds at 25 km.}
\label{tab:noise_boundaries}
\centering
\small
\begin{tabularx}{\columnwidth}{@{}>{\raggedright\arraybackslash}Xrrr@{}}
\toprule
Boundary & $\xi_{\rm sh}$ & Penalty & SCR \\
 & (mSNU) & (dB) & (Mbit/s) \\
\midrule
Shared-versus-separated hardware threshold & 28.718 & 1.761 & 8.064 \\
Zero-key-rate threshold & 105.844 & $\infty$ & $0$ \\
\bottomrule
\end{tabularx}
\end{table}

Under the nominal transmitter spectrum, the sharing penalty reaches $0.444~\mathrm{dB}$ over $25$--$100~\mathrm{km}$. This is below the $1.761$-dB threshold for the modeled one-direction resource index. A Raman-coefficient sweep over $(1.5$--$3.1)\times10^{-9}~\mathrm{km^{-1}nm^{-1}}$ gives $100$-km penalties from $0.221$ to $0.459~\mathrm{dB}$.

Each row in Table~\ref{tab:transmitter_noise_sensitivity} is an independent sensitivity scenario in which the integrated admitted fraction $\epsilon_{\rm OOB}$, defined before polarization suppression in Eq.~\eqref{eq:oob_fraction}, is prescribed and all forward-link variables are reoptimized. Because the receive filter and signal placement change between designs, evaluating a particular transmitter requires integration of its measured spectrum at each candidate operating point. At $10\log_{10}\epsilon_{\rm OOB}=-60~\mathrm{dB}$, the sharing penalty is $0.026~\mathrm{dB}$ at $25~\mathrm{km}$ and $2.058~\mathrm{dB}$ at $100~\mathrm{km}$, exceeding the adopted $1.761$-dB resource-index threshold at the longer distance.

\begin{table}[!t]
\caption{Reoptimized transmitter-noise sensitivity. The nominal case has $\epsilon_{\rm OOB}=0$. Rates are asymptotic design projections; zero denotes no positive rate found within the search bounds.}
\label{tab:transmitter_noise_sensitivity}
\centering\scriptsize
\setlength{\tabcolsep}{3pt}
\begin{tabular}{@{}lrrrr@{}}
\toprule
$10\log_{10}\epsilon_{\rm OOB}$ & \multicolumn{2}{c}{25 km} & \multicolumn{2}{c}{100 km}\\
(dB) & SCR & Penalty & SCR & Penalty\\
 & (Mbit/s) & (dB) & (Mbit/s) & (dB)\\
\midrule
Nominal & 12.079 & 0.006 & 0.205 & 0.444 \\
-80 & 12.078 & 0.006 & 0.204 & 0.458 \\
-70 & 12.073 & 0.008 & 0.198 & 0.586 \\
-60 & 12.022 & 0.026 & 0.141 & 2.058 \\
-50 & 11.528 & 0.209 & 0.000 & $\infty$ \\
-40 & 7.473 & 2.091 & 0.000 & $\infty$ \\
\bottomrule
\end{tabular}
\end{table}

The transport-efficiency sweep changes the nominal shared SCR over $\eta_c\in[0.90,1.00]$ from $12.078$ to $12.079~\mathrm{Mbit/s}$ at $25~\mathrm{km}$, and from $0.204$ to $0.206~\mathrm{Mbit/s}$ at $100~\mathrm{km}$.

\section{Discussion}
\label{sec:discussion}

The central result is that sharing the transmitter and fiber within one modeled direction can preserve most of the separated-link service rate at metropolitan distance. At the selected $25$-km design, the nominal sharing-induced excess noise is $0.101~\mathrm{mSNU}$, approximately $0.7\%$ of the baseline input-referred excess noise. Adequate spectral separation therefore makes the sharing penalty small within the adopted transmitter model.

With return capacity provisioned independently, the Alice-to-Bob classical channel carries parameter-estimation data and the OTP payload, which set its required symbol rate and minimum launch power. At the selected $25$-km design, parameter-estimation traffic is approximately $13.4$ times the OTP payload rate; the resulting classical rate and launch-power requirements influence the key rate through spectral leakage and Raman scattering.

Spectral separation reduces leakage from the intended classical waveform, while residual transmitter noise within the quantum receive band contributes according to the power admitted by the receive filter. Table~\ref{tab:transmitter_noise_sensitivity} shows that the longer-distance link tolerates less additional transmitter noise. At the nominal operating points in Table~\ref{tab:distance_points}, $P_c$ rises from $0.0059$ to $0.1747~\mathrm{mW}$ between $25$ and $100~\mathrm{km}$, while the quantum symbol rate and polarization suppression remain fixed; the same additional admitted fraction $\epsilon_{\rm OOB}$ would therefore contribute roughly $30$ times more input-referred excess noise at $100~\mathrm{km}$, according to Eq.~\eqref{eq:leakage_input_referred}.

The baseline excess noise $\xi_A$ is inferred from data obtained after the
complete receiver DSP and is then referred to the channel input. It therefore
includes the residual phase-recovery noise of the characterized configuration
\cite{RicardThesis2025}. A future architecture-specific model can separate this
contribution by comparing the residual phase-error variance of
classical-assisted phase recovery in the shared link with that of the
phase-reference method adopted for the separated link.

An experimental assessment of these projections would combine measurements of the transmitter spectrum, the polarization extinction ratio, and the absolute launch-power reference with the receiver frequency response, electronic- and shot-noise spectra, and DSP efficiency. Calibration at the selected channel separations and symbol rates would establish the Raman coefficient and classical SNR scaling for the platform, while coding tests would assess reliable authenticated transport within the provisioned capacities.

\section{Conclusion}
\label{sec:conclusion}

A classical/DM-CV-QKD link with one shared transmitter and fiber per modeled transmission direction has been formulated as a coupled optimization of security, transport, spectrum, launch power, and hardware resources. The framework links the binomial 64-QAM security calculation and trusted receiver model to the classical traffic required by the QKD service and to the leakage and Raman noise introduced by sharing.

The optimized shared Alice-to-Bob link retains nearly the separated-link service rate at metropolitan distance while reducing duplicated transmitter and fiber resources in that direction. The reported rates use the independently provisioned return capacity of the one-direction model. Under the nominal transmitter spectrum, the illustrative
bandwidth--noise scenarios give an optimized $25$-km one-direction
SCR of $12.079~\mathrm{Mbit/s}$ at $350~\mathrm{MHz}$ and
$24.535~\mathrm{Mbit/s}$ at $750~\mathrm{MHz}$. At longer distances, the sharing benefit becomes more sensitive to additional transmitter noise in the quantum receive band.

The next stage is to apply the same co-design framework with a composable finite-size DM-CV-QKD security model and calibration data for further receiver/DSP integration, extending the resource reduction toward the fully integrated architecture in Fig.~\ref{fig:resource_sharing_vision}.

\section*{Code Availability}

The Python code used for the numerical simulations is available in the public GitHub repository\footnote{\url{https://github.com/shivang-sri-vastava/SCR-Optimization-QKD}}. The repository README provides the software requirements and instructions for running the simulations.

\appendices

\section{Security-Model Implementation Details}
\label{app:security_backend}

This appendix gives the operator-level definitions used to evaluate the binomial 64-QAM covariance bound and trusted-receiver purification in Section~\ref{subsec:quantum_rate_model}. Let $a$ denote the bosonic annihilation operator. The average modulation state and support-restricted operator are
\begin{equation}
 \tau=\sum_{k,\ell}p_{k\ell}|\alpha_{k\ell}\rangle\langle\alpha_{k\ell}|,
 \qquad
 a_\tau=\tau^{1/2}a\tau^{-1/2},
 \label{eq:tau_operator}
\end{equation}
where the inverse is restricted to the support of $\tau$. The constellation quantities entering Eq.~\eqref{eq:zstar_gaussian_channel} are \cite{Denys2021}
\begin{align}
 C_\tau&=\mathrm{Tr}\!\left(\tau^{1/2}a\tau^{1/2}a^\dagger\right),
 \label{eq:Ctau}\\
 w&=\sum_{k,\ell}p_{k\ell}\!\left[
 \langle a_\tau^\dagger a_\tau\rangle_{k\ell}
 -|\langle a_\tau\rangle_{k\ell}|^2\right].
 \label{eq:w_constellation}
\end{align}

For a phase-insensitive channel, $\xi_{\rm link}$ from
Eq.~\eqref{eq:link_excess_noise} is the input-referred link contribution. The
analytical correlation bound and the pre-receiver covariance are
\begin{align}
 Z^\star
 &=2\sqrt{T_{\rm ch}}C_\tau
 -\sqrt{2T_{\rm ch}\xi_{\rm link}w},
 \label{eq:zstar_gaussian_channel}\\
 \Gamma_{AB_1}
 &=\begin{pmatrix}
 V\mathbf I_2 & Z^\star\boldsymbol{\sigma}_z\\
 Z^\star\boldsymbol{\sigma}_z & W\mathbf I_2
 \end{pmatrix},
 \label{eq:pre_receiver_covariance}
\end{align}
where $V=V_A+1$, $W=1+T_{\rm ch}(V_A+\xi_{\rm link})$,
$\mathbf I_2$ is the $2\times2$ identity matrix,
$\boldsymbol{\sigma}_z=\mathrm{diag}(1,-1)$, and $B_1$ denotes Bob's mode before
trusted detection. The receiver contribution is separated before constructing
the covariance assigned to Eve.

Following the detector-output variance convention in \cite{RicardThesis2025},
Eq.~\eqref{eq:detected_noise_budget} converts the channel-input-referred
contribution to each detected heterodyne quadrature. The factor $1/2$ follows
from heterodyne detection, and the separately identified terms are the shot
noise $N_0$, trusted electronic noise $\xi_{\rm rec}$, and link excess noise
$\xi_{\rm link}$.

The trusted-receiver purification variance is \cite{AymericThesis2022,RoumestanThesis}
\begin{equation}
 W_{\rm rec}=1+\frac{2\xi_{\rm rec}}{1-\eta_B},
 \label{eq:trusted_receiver_variance}
\end{equation}
and the corresponding Einstein--Podolsky--Rosen (EPR) covariance of the input
ancilla modes $F'$ and $G$ is
\begin{equation}
 \Gamma_{F'G}=\begin{pmatrix}
 W_{\rm rec}\mathbf I_2 & c_{\rm rec}\boldsymbol{\sigma}_z\\
 c_{\rm rec}\boldsymbol{\sigma}_z & W_{\rm rec}\mathbf I_2
 \end{pmatrix},
 \quad c_{\rm rec}=\sqrt{W_{\rm rec}^2-1}.
 \label{eq:trusted_receiver_epr}
\end{equation}
Let $B_{B_1F'}(\eta_B)$ denote the beam-splitter symplectic transformation mixing $B_1$ with the receiver ancilla $F'$. Applying it to $\Gamma_{AB_1}\oplus\Gamma_{F'G}$ and reordering the modes as $(A,F,G,B)$ gives
\begin{equation}
 \Gamma_{AFGB}
 =\begin{pmatrix}
 \Gamma_{AFG} & \Gamma_C\\
 \Gamma_C^T & \Gamma_B
 \end{pmatrix}.
 \label{eq:trusted_receiver_partition}
\end{equation}
Conditioning on Bob's heterodyne result gives
\begin{equation}
 \Gamma_{AFG|Y}=\Gamma_{AFG}
 -\Gamma_C(\Gamma_B+\mathbf I_2)^{-1}\Gamma_C^T.
 \label{eq:conditional_trusted_covariance}
\end{equation}
For an $n$-mode covariance matrix $\Gamma$, its symplectic eigenvalues are the positive moduli of the eigenvalues of $i\Omega_n\Gamma$, where $\Omega_n=\bigoplus_{j=1}^n\left(\begin{smallmatrix}0&1\\-1&0\end{smallmatrix}\right)$. If $\nu_{1,2}$ belong to $\Gamma_{AB_1}$ and $\nu_{3,4,5}$ belong to Eq.~\eqref{eq:conditional_trusted_covariance}, then
\begin{equation}
 \chi(E;Y)=\sum_{i=1}^{2}g\!\left(\frac{\nu_i-1}{2}\right)
 -\sum_{i=3}^{5}g\!\left(\frac{\nu_i-1}{2}\right),
 \label{eq:trusted_holevo}
\end{equation}
where $g(x)=(x+1)\log_2(x+1)-x\log_2x$. One conditional eigenvalue is unity and contributes zero entropy.

\section{Numerical Validation}
\label{app:numerical_validation}

The discrete-modulation calculation uses a Fock-space cutoff of $60$, and the mutual-information integral uses $32$ Gauss--Hermite nodes per quadrature. The constellation quantities and mutual information are evaluated on dense grids and interpolated with shape-preserving piecewise cubic Hermite polynomials during the repeated optimization calls. Direct evaluations are retained for convergence checks. Increasing both numerical orders changes the separated-link reference rate by less than $2\times10^{-8}\%$. All covariance matrices reported in the design sweeps satisfy $\nu_i\geq1$ within numerical tolerance, and the trusted-receiver calculation converges to the ideal heterodyne conditional entropy when $\eta_B\rightarrow1$ and $\xi_{\rm rec}\rightarrow0$.

At the $25$-km separated-link reference of Eq.~\eqref{eq:separated_reference_result}, the symplectic eigenvalues are
\begin{equation}
\begin{aligned}
 (\nu_1,\nu_2)&=(1.0087,4.4330),\\
 (\nu_3,\nu_4,\nu_5)&=(1.0000,1.0045,3.6743).
\end{aligned}
 \label{eq:reference_symplectic_values}
\end{equation}
The unit conditional eigenvalue is consistent with the receiver purification used in Eq.~\eqref{eq:trusted_holevo}.

\EOD
\end{document}